\documentclass{aa}  
 
\usepackage{graphicx}
\usepackage{txfonts}
\usepackage{hyperref}
\hypersetup{colorlinks,allcolors=blue}

\usepackage{subcaption} 
\usepackage{amssymb} 
\usepackage[dvipsnames]{xcolor} 
\usepackage{ tipa } 
\providecommand{\orden}[1]{\ensuremath{10^{#1}}} 
\usepackage{siunitx} 
\usepackage[flushleft]{threeparttable}

\newcommand{\qhired}{$\chi^{2}_{\nu}$}

\newcommand{\Msun}{M$_\mathrm{\odot}$}%modificado
\newcommand{\MNS}{$M_\mathrm{NS}$} %nuevo
\newcommand{\ergs}{erg~s$^{-1}$}

\newcommand{\kms}{\mbox{km~s$^{-1}$}}

\usepackage{fancyhdr}
\usepackage{natbib}

\begin{document} 

   \title{Optical nebular emission following the most luminous outburst of Aquila~X-1}
    \author{G. Panizo-Espinar \inst{1}\textsuperscript{,}\inst{2}\fnmsep\thanks{E-mail: guayente.panizo@gmail.es}
       \and T. Muñoz-Darias \inst{1}\textsuperscript{,}\inst{2} 
       \and M. Armas Padilla\inst{1}\textsuperscript{,}\inst{2}
       \and F. Jiménez-Ibarra\inst{1}\textsuperscript{,}\inst{2}
       \and J. Casares\inst{1}\textsuperscript{,}\inst{2}
       \and D.~Mata~S\'anchez\inst{3}
       }
    \authorrunning{G. Panizo-Espinar et al.} %Short list for the running heads
    \titlerunning{Optical nebular emission from Aquila~X-1}
    \institute{Instituto de Astrofísica de Canarias (IAC), Vía Láctea, La         Laguna, E-38205, Santa Cruz de Tenerife, Spain
        \and
        Departamento de Astrofísica, Universidad de La Laguna, E-38206 Santa Cruz de Tenerife, Spain 
        \and
        Jodrell Bank Centre for Astrophysics, Department of Physics and Astronomy, The University of Manchester, Manchester M13 9PL, UK
    }
   \date{Received 12 January 2021; accepted 31 March 2021}

% \abstract{}{}{}{}{} % 5 {} token are mandatory
  \abstract {Aquila X-1 is a prototypical neutron star low mass X-ray binary and one of the most studied X-ray transients. We present optical spectroscopy obtained with the \textit{Gran Telescopio Canarias} (10.4 m) during the 2016 outburst, the brightest recorded in recent times and which showed a standard evolution with hard and soft accretion states. Our dataset includes a dense coverage of the brightest phases of the event, as well as the decay towards quiescence. We searched for optical winds by studying the profiles and evolution of the main emission lines and found no indisputable wind signatures, such as P-Cyg profiles. Nonetheless, our detailed analysis of the particularly strong and broad H$\alpha$ emission line, detected at the end of the outburst, is consistent with the presence of a nebular phase produced by optically thin ejecta at $\sim$~800 \kms \ or, alternatively, an extended disc atmosphere. We discuss these possibilities as well as the similarities with the phenomenology observed in other black hole and neutron star systems. Our study suggests that optical nebular phases might be a relatively common observational feature during the late stages of low mass X-ray binaries' outbursts, enabling us to probe the presence of outflows at low-to-intermediate orbital inclinations.}

   \maketitle
%%%%%%%%%%%%%%%%%%%%%%%%%%%%%%%%%%%%%%%%%%%%%%%%%%%%%%%%%%%%%%%%%%%%%%
% INTRODUCTION
%%%%%%%%%%%%%%%%%%%%%%%%%%%%%%%%%%%%%%%%%%%%%%%%%%%%%%%%%%%%%%%%%%%%%%
    \section{Introduction}\label{intro}
    
    Low mass X-ray binaries (LMXBs) are composed of a stellar-mass black hole (BH) or a neutron star (NS) that accretes  material from a low mass donor ($\lesssim$ 1 \Msun) via Roche lobe overflow. Since the angular momentum must be conserved, the infalling material creates an accretion disc around the compact object \citep{ShakuraSunyaev1973}, whose innermost areas reach temperatures of $\sim$10$^{7}$ K, and therefore radiate in X-rays. 

        Some LMXBs are always active and emit large amounts of X-ray radiation persistently. However, most of them are transient sources that go through two different activity phases. They spend most of their lifespans in a quiescent state, showing occasional episodes of enhanced accretion in which their optical and X-ray luminosities increase by several orders of magnitude. These episodes are so-called outbursts, and they can last from weeks to years (see e.g. \citealt*{Casares2017}; \citealt{Corral-Santana2016, Tetarenko2016}).
        During an outburst, the X-ray spectrum evolves according to the properties of the accretion flow. It can be dominated by either a soft, thermal component, arising in the accretion disc, or by a hard one with a power-law shape, thought to be produced by inverse-Compton processes in a corona of hot electrons (e.g. \citealt{Gilfanov2010}).  Depending on which component dominates the X-ray emission, the system can be found in soft, hard, or intermediate states (\citealt{McClintock2006}; \citealt*{ Done2007,Belloni2011}). The spectral analysis is significantly more complex in NS systems than in BHs due to the presence of an additional thermal component arising in the NS surface (e.g. \citealt{vanDerKlis2006,Lin2007,Lin2009,ArmasPadilla2017,ArmasPadilla2018, Burke2017}). 
        
        Black hole transients typically follow a canonical evolution through the different states, displaying anticlockwise loop patterns in the hardness-intensity diagram (HID, \citealt{Homan2001}). During the initial rise (throughout the hard state), the X-ray luminosity increases by several orders of magnitude (from  $\sim$\orden{31} to $\sim$ \orden{35-37} \ergs ). Then, a fast transition to the soft state is observed, followed by a much slower decay in luminosity. Finally, the system returns to the hard state through a different track in luminosity, drawing a hysteresis pattern (see e.g. \citealt{Fender2012}). These hysteresis loops are found when studying the evolution of other observables, such as the fast variability (\citealt{Munoz-Darias2011}, \citealt{Heil2012}), and are also a common feature in NS LMXBs (both transient and persistent) when accreting at intermediate rates (\citealt{Munoz-Darias2014}; see also \citealt{Maccarone2003}).

        The above-described accretion processes are tightly coupled to multi-wavelength outflow phenomena (e.g. \citealt{Fender2016}). In BH LMXBs, radio emission from a compact, unresolved jet is universally observed in the hard state, but not detected during the soft state (\citealt{Fender2004}; see also \citealt{Russell2011}). Contrastingly, hot accretion disc X-ray winds are typically observed during soft states, with terminal velocities of $\sim$100--2000~\kms (\citealt{Neilsen2009, Ponti2012}). BH transients also show optical disc winds that can be strictly simultaneous with the radio jet, producing P-Cygni profiles in recombination lines of helium and hydrogen. Unusually broad emission components have also been witnessed, which are sometimes simultaneous with P-Cyg profiles (e.g. fig. 15 in \citealt{MataSanchez2018}) and linked to the presence of winds. These cold optical outflows have been observed so far in several BH transients (see e.g. \citealt{Casares1991, Munoz-Darias2016,Munoz-Darias2017,Munoz-Darias2018,Munoz-Darias2019,Charles2019, JimenezIbarra2019b, Cuneo2020b}), but they have not yet been observed in soft states, which might be related to ionisation or other wind-visibility effects. This scenario is supported by the detection of near-infrared wind signatures in a BH soft state that does not show optical wind features (\citealt{Sanchez-Sierras2020}). LMXBs with NS accretors largely share this complex accretion-ejection coupling, showing jets (\citealt{MigliariFender2006},  \citealt{MillerJones2010}) and hot winds  (\citealt{Ponti2014, Ponti2015, DiazTrigo2016}). Near-infrared P-Cyg profiles have also been observed in high-luminosity NS systems (\citealt{Bandyopadhyay1999}), and, more recently, conspicuous optical wind signatures were discovered in Swift~J1858.6-0814 (\citealt{Munoz-Darias2020}), a NS transient with a suspected orbital period of $\sim$21 hr (\citealt{Buisson2020b, Buisson2020c}).   
        
    In addition, luminous BH outbursts approaching or occasionally exceeding the Eddington limit (\citealt{Revnivtsev2002,Motta2017}) have shown unprecedentedly strong optical emission lines with broad wings following particularly sharp outburst declines (\citealt{Munoz-Darias2016, Munoz-Darias2018}). These prominent lines are thought to be produced during the optically thin expanding phase of previously launched ejecta, which would cool down and recombine as the central source flux drops. This is the so-called nebular phase (\citealt{Munoz-Darias2016, Casares2019}; see also  \citealt{Rahoui2017}) that was first noticed in the final stages of the 2015 outburst of V404 Cyg. However, less pronounced versions of this phenomenology were also witnessed in earlier phases of this event \citep{MataSanchez2018}, as well as in other BH transients. This is the case of MAXI~J1820+070, with a peak luminosity of (only) 15\% L$_\mathrm{Edd}$ (\citealt{Atri2020}), that showed prominent broad emission line wings and other wind-related features during the hard state peak (\citealt{Munoz-Darias2019}).

        In this paper, we present the results of a spectroscopic search for cold optical winds in Aquila X-1 (hereafter Aql~X-1) during its 2016 outburst, the most energetic in recent times with a peak luminosity of $\sim$~0.5 L$_\mathrm{Edd}$ (\citealt{Gungor2017}). This NS transient is one of the most intensively studied LMXBs thanks to its short recurrence outburst period of $\sim$~0.8 years \citep*{Campana2013}. The system usually brightens from magnitude $V=21.6$ in quiescence (\citealt{Chevalier1999}) to $V\sim15-17$ in outburst \citep{Garcia1999}. Aql~X-1 is seen through a relatively low orbital inclination ($i$), $23^\circ$< $i$ <$53^\circ$ (assuming \MNS \ < 3\Msun, \citealt{MataSanchez2017}) and possesses one of the longest orbital periods in a NS transient (18.72~h; \citealt{Chevalier1991,MataSanchez2017}).

% ------------------------ Figure 1 --------------------------
    \begin{figure}
    \centering
    \begin{subfigure}{\columnwidth}
        \includegraphics[width=9truecm]{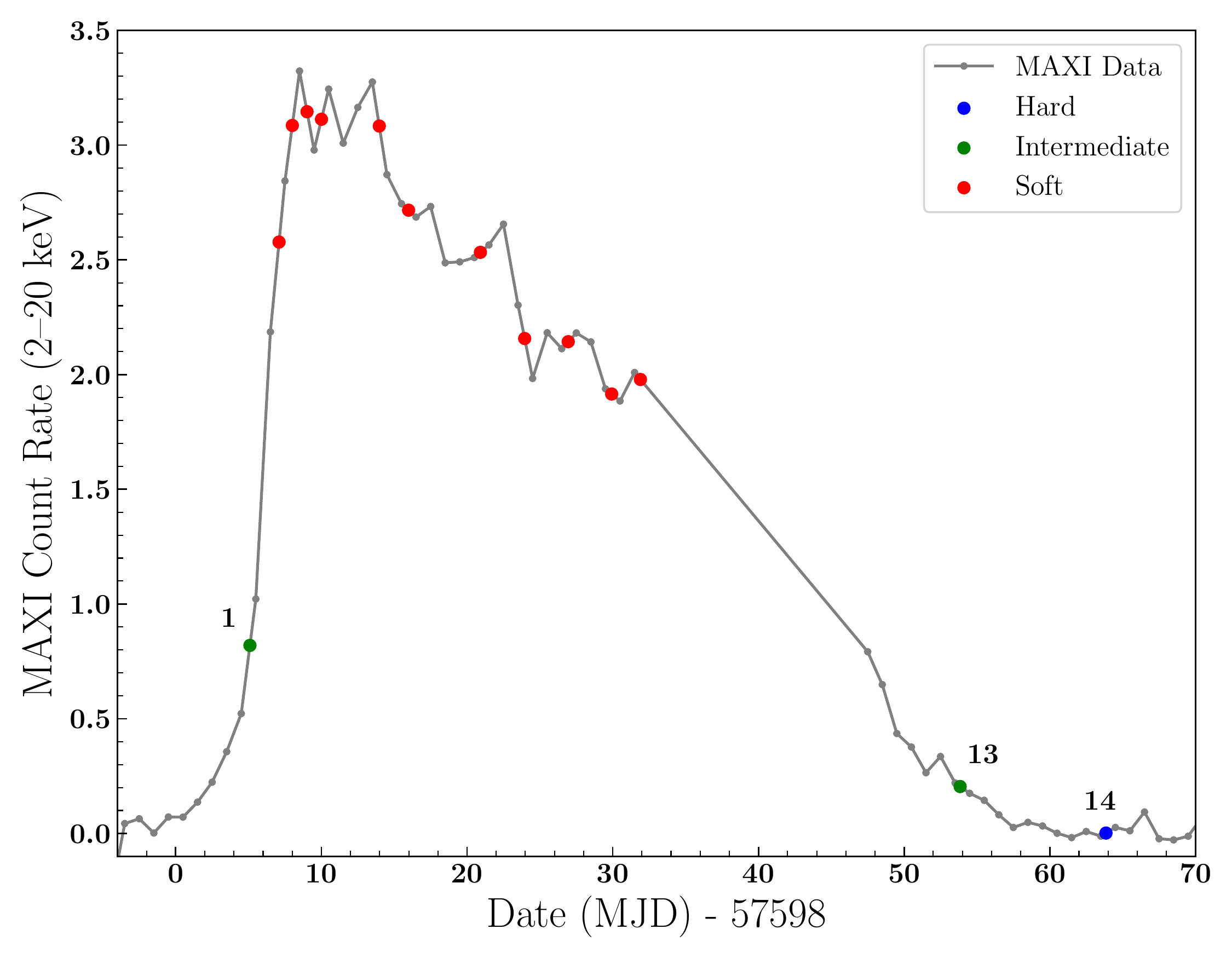}
    \end{subfigure}
    \begin{subfigure}{\columnwidth}
        \includegraphics[width=9truecm]{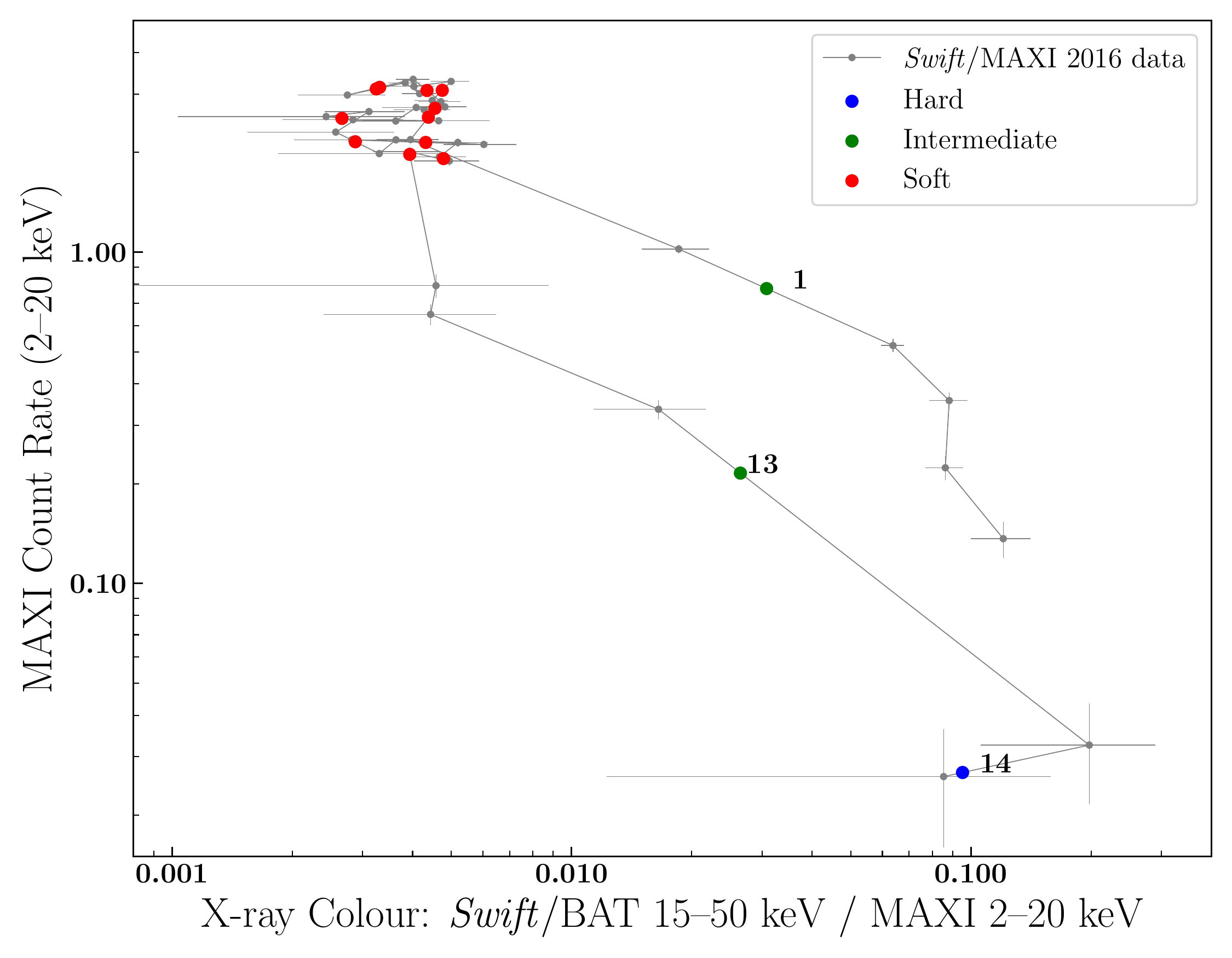}
    \end{subfigure}
    \caption{Light curve and HID of Aql~X-1 during its 2016 outburst. Grey dots represent the X-ray data, while the larger dots (coloured according to the X-ray state; see legend), mark the interpolated position for each GTC epoch. Epochs \#1, \#13, and \#14 are marked with numbers.  \textit{Upper pannel:} Light curve from MAXI
   (2--20 keV). \textit{Bottom pannel:} HID obtained by using \textit{Swift}/BAT (15--50 keV; hard band) and MAXI (2--20 keV; soft band), starting on MJD 57598. We only considered count rates exceeding 0.0022 photons cm$^{-2}$ s$^{-1}$ (MAXI) and 0.01 counts cm$^{-2}$ s$^{-1}$ (Swift/BAT). The source displays the usual hysteresis loop in the anticlockwise direction.}
    \label{fig1_light_HID}
    \end{figure}

        \begin{table*}
        \centering
        \caption{Observing log and corresponding X-ray state.}
        \label{Table_data}
        \begin{threeparttable}
        \begin{tabular}{c c c c c}
                \hline\hline
                Epoch & 2016 Date (MJD) & Accretion state & Grism\tnote{a} \ and exposures & g-band magnitude (\# of observations)\\
                \hline
                1 & 03 Aug (57603)      & Intermediate   & R1 (1$\times$300s) + R2 (6$\times$700s) & 17.80 $\pm$ 0.02 (2) \\ 
                2 & 05 Aug (57605)      & Soft  & R1 (1$\times$300s) + R2 (8$\times$600s) & 16.66 $\pm$ 0.02 \\
                3 & 06 Aug (57606)      & Soft  & R1 (1$\times$300s) + R2 (2$\times$600s) & 16.72 $\pm$ 0.02\\
                4 & 07 Aug (57607)  & Soft  & R1 (1$\times$300s) + R2 (2$\times$600s) & 16.49 $\pm$ 0.02 \\
                5 & 08 Aug (57608)      & Soft  & R1 (1$\times$300s) + R2 (2$\times$600s) & 16.50 $\pm$ 0.02\\
                6 & 11 Aug (57611)      & Soft  & R1 (1$\times$300s) + R2 (2$\times$600s) & 16.08 $\pm$ 0.02 (2) \\
                7 & 13 Aug (57613)      & Soft  & R1 (2$\times$300s) + R2 (2$\times$600s) & 16.54 $\pm$ 0.04 (6) \\
                8 & 18 Aug (57618)      & Soft  & R1 (2$\times$150s) + R2 (4$\times$300s) & 16.65 $\pm$ 0.02\\
                9 & 21 Aug (57621)      & Soft  & R1 (1$\times$300s) + R2 (2$\times$600s) & 16.181 $\pm$ 0.004\tnote{*} \\
                10 & 24 Aug (57624)     & Soft  & R1 (1$\times$300s) + R2 (2$\times$600s) & 16.48 $\pm$ 0.02 (3) \\
                11 & 27 Aug (57627) & Soft  & R1 (1$\times$300s) + R2 (2$\times$600s) & 16.53 $\pm$ 0.02 (3) \\
                12 & 29 Aug (57629)     & Soft  & R1 (1$\times$300s) + R2 (3$\times$685s) & 16.51 $\pm$ 0.02 (3) \\
                13 & 20 Sep (57651)     & Intermediate   & R1 (2$\times$400s) & 18.90 $\pm$ 0.02\\
                14 & 30 Sep (57661)     & Hard  & R1 (2$\times$400s) &  19.82 $\pm$ 0.03\tnote{**} (2) \\
                \hline
        \end{tabular}
    \begin{tablenotes}
    \item[a] R1 and R2 indicate R1000B and R2500V grisms, respectively.
    \item[*] The acquisition image in this night was obtained in the r-band, so we report this value instead.
    \item[**] Aql~X-1 has an interloper star at $\sim$0.5\textquotedbl \ \citep{Chevalier1997, MataSanchez2017}. This magnitude is consistent with a significant contribution from the interloper to the optical spectrum (see Sect. \ref{interloper}).
    \end{tablenotes}
    \end{threeparttable}
        \end{table*}
        
%%%%%%%%%%%%%%%%%%%%%%%%%%%%%%%%%%%%%%%%%%%%%%%%%%%%%%%%%%%%%%%%%%%%%%
% OBSERVATIONS
%%%%%%%%%%%%%%%%%%%%%%%%%%%%%%%%%%%%%%%%%%%%%%%%%%%%%%%%%%%%%%%%%%%%%%
    \section{Observations and data reduction}\label{observations}
    A total of 55 spectra (Table~\ref{Table_data}) were obtained during August and September 2016 using the Optical System for Imaging and low-Intermediate-Resolution Integrated Spectroscopy (OSIRIS, \citealt{Cepa2000}) at the Gran Telescopio Canarias (GTC) in the Observatorio del Roque de los Muchachos (La Palma, Spain). 
    We used the R1000B grism (2.12~\mbox{\normalfont\AA} pix$^{-1}$), which covers the 3630--7500~\mbox{\normalfont\AA} spectral range, and R2500V (0.8~\mbox{\normalfont\AA} pix$^{-1}$), covering the range of 4500--6000~\mbox{\normalfont\AA}. We used a slit width of 0.8 arcsec (1 arcsec for epoch \#1), which rendered a velocity resolution of $\sim$330 and $\sim$130~\kms \  for the R1000B and R2500V grisms, respectively ($\sim$370 and $\sim$170~\kms \  for epoch \#1). During the observing campaign, the seeing was between 0.8 and 1.4 arcsec depending on the epoch. Observations were slit-limited, except for epoch \#1.
    
    \subsection{Data reduction}

    Data were reduced using \textsc{iraf}\footnote{\textsc{iraf} is distributed by the National Optical Astronomy Observatory, which is operated by the Association of Universities for Research in Astronomy, Inc. under contract to the National Science Foundation.} standard routines. Regular HgAr+Ne and HgAr+Ne+Xe arc lamp exposures taken on each observing block were used to perform the pixel-to-wavelength calibration. This was corrected from flexure effects (<~115~\kms) using the \textsc{molly} software and the \ion{O}{i} 5577.34~\mbox{\normalfont\AA} \ sky line. Spectra were daily averaged in 14 epochs and normalised. No higher resolution (R2500V) data were taken in epochs \#13 and \#14 (see Table \ref{Table_data}). Finally, we subtracted the systemic velocity of 104~\kms \ reported in \citet{MataSanchez2017} to the spectra.
    
    Flux calibration of the R1000B data was performed against ESO\footnote{Standard stars calibration files from ESO: ftp://ftp.eso.org/pub/stecf/standards/okestan/, accessed 6 July 2020.} and ING\footnote{Calibration fields from Isaac Newton Group of Telescopes: http://catserver.ing.iac.es/landscape/tn065-100/workflux.php.} Spectophotometric Standards \citep{Oke1974, Oke1990} depending on the epoch (Feige 66, Feige 110, G191-B2B, Ross 640 and G158-100) and using \textsc{astrophy-photutils}-based routines \citep{Bradley2019}. We also applied a reddening correction of E$_{B-V}$ = 0.65 mag (\citealt{LopezNavas2020}). The main aim of this task is not to report absolute flux measurements but to compute line intensity ratios. We note that this calibration is not as accurate as, for example, that obtained by comparing the target with a field star included in the slit, but it is good enough for our purposes.
    
    We also obtained photometric data (\textit{g} band) from the acquisition images preceding each spectroscopic observation. They were reduced using \textsc{astropy-ccdproc}-based routines \citep{Astropy2013} and calibrated against nearby stars present in the PanSTARRs catalogue. Averaged magnitudes are reported in Table \ref{Table_data}. 
    
\subsection{Possible contamination by the interloper star}
\label{interloper}

A G8V interloper star (\textit{V}=19.42 $\pm$ 0.06) located 0.48" east of Aql X-1 dominates the combined optical flux during quiescence \citep{Chevalier1999}, particularly in the blue part of the spectrum. This might also be relevant for our faintest epochs, which are \#13 and \#14 with \textit{V}$\sim$18 and \textit{V}$\sim$19.4, respectively (see fig.1 in \citealt{DiazTrigo2018}). Consistent numbers are obtained by comparing our \textit{g}-band magnitudes (\textit{g}=18.90 and \textit{g}=19.82; see Table \ref{Table_data}) with the minimum flux (i.e. in quiescence) recorded by the PANSTARRS survey (\textit{g}=19.95; \citealt{Chambers2016}). This indicates that a very significant, if not dominant, contribution from the interloper is expected in epoch \#14. However, in epoch \#13 Aql~X-1 is still 2.5 times brighter than the interloper in the V band. Furthermore, the 0.8" slit was oriented along the south-north direction, and therefore at least 50\% of its flux was left out. Taking into account the distance to the interloper, the slit width and the seeing in epoch \#13, we estimate an interloper's contribution of $\sim$12\% in the \textit{V}-band ($\sim$15\% in the \textit{g}-band).

% ------------------------ Figura 2  --------------------------
    \begin{figure*}
    \centering
    \begin{subfigure}{2\columnwidth}
        \includegraphics[width=17.5truecm]{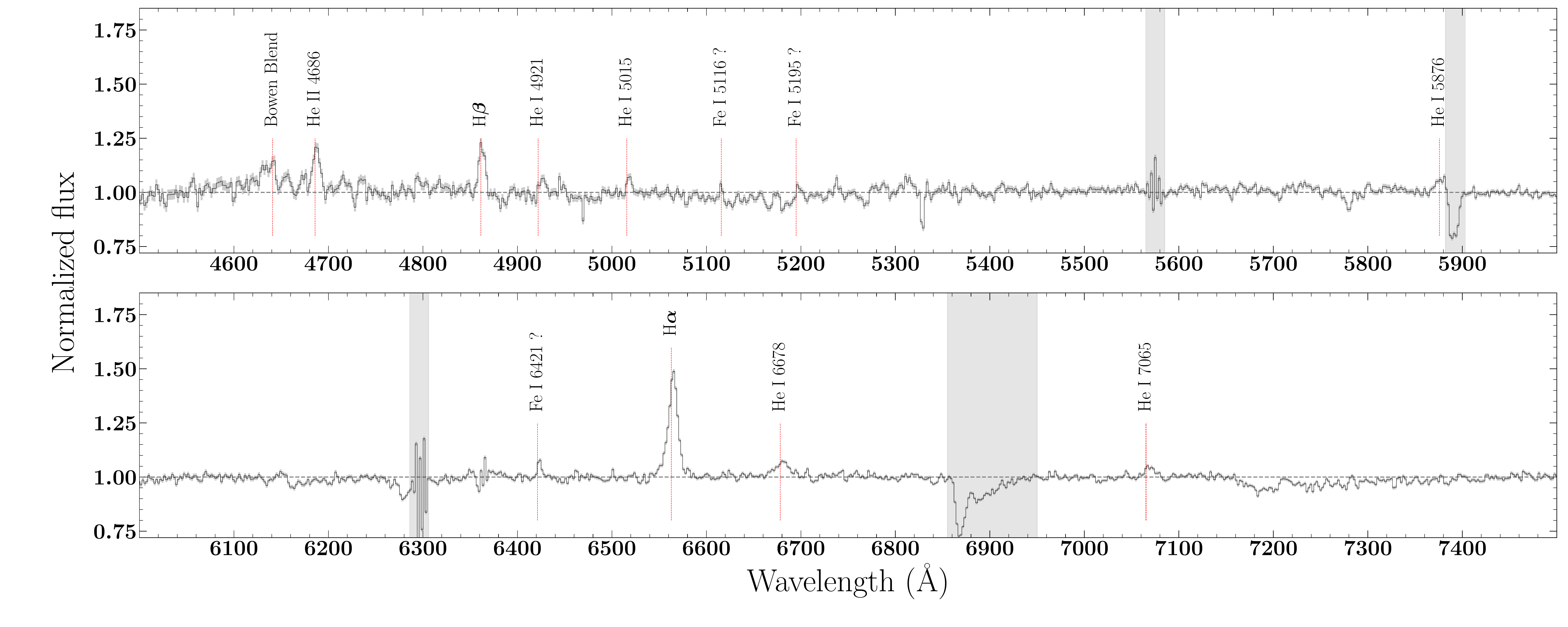}
    \end{subfigure}
    \caption{Example of normalised spectrum (epoch \#13). The most prominent spectral lines are indicated with red lines. Grey-shaded regions mark telluric absorptions and residuals from the sky subtraction.}
    \label{fig2new}
    \end{figure*}       
% --------------------------------------------------------

% ------------------------ Figure 3 --------------------------
\begin{figure}
\centering
    \begin{subfigure}{\columnwidth}
         \includegraphics[width=9truecm]{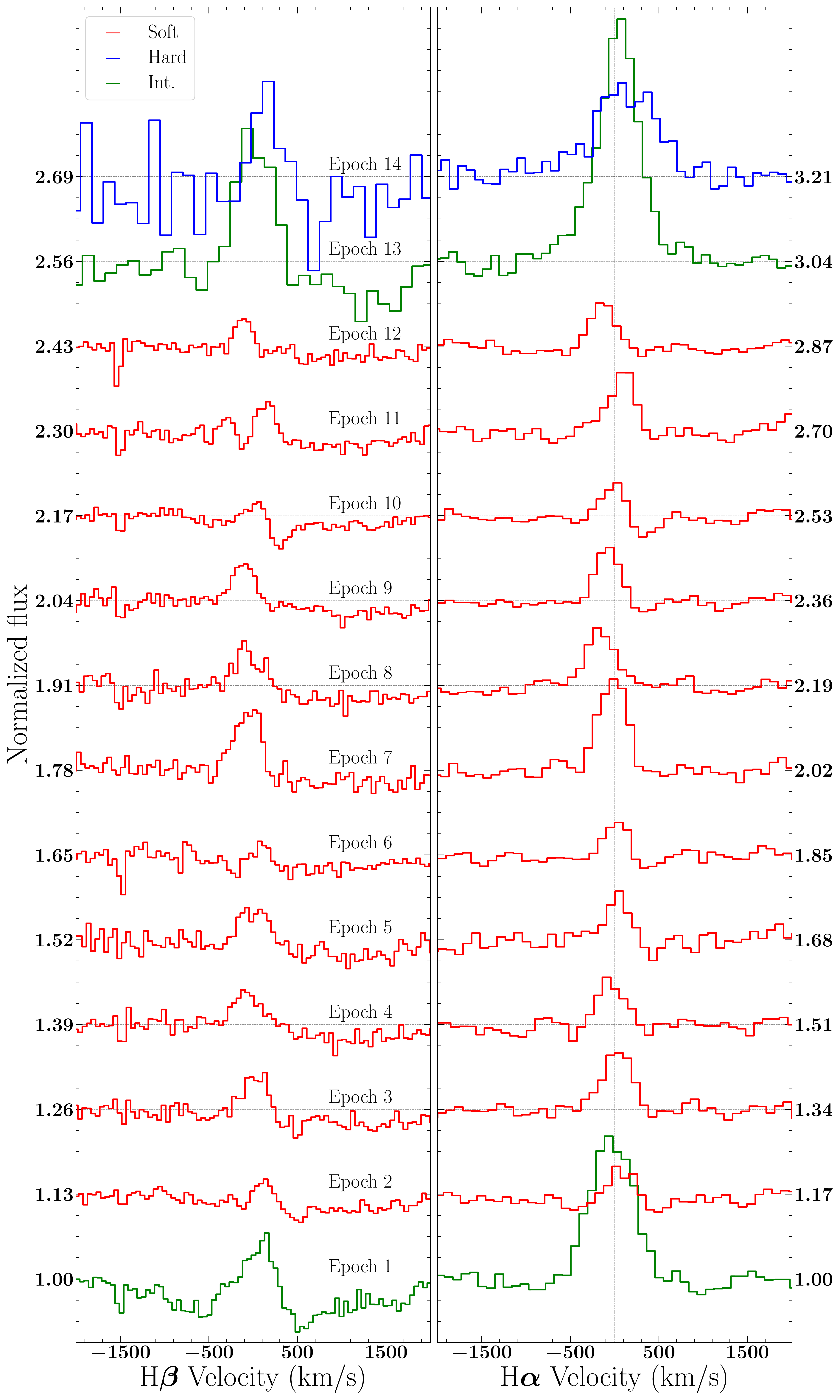}
    \end{subfigure}
    \caption{Temporal evolution of H$\beta$ (left) and H$\alpha$ (right). Offsets of 0.13 (H$\beta$) and 0.17 (H$\alpha$) were used. Colour code is the same as in Fig. \ref{fig1_light_HID}.}
    \label{fig2_evolution}
\end{figure}
% --------------------------------------------------------

 % ------------------------ Figure 4 --------------------------
\begin{figure*}
\centering
    \begin{subfigure}{\columnwidth}
        \includegraphics[width=9truecm]{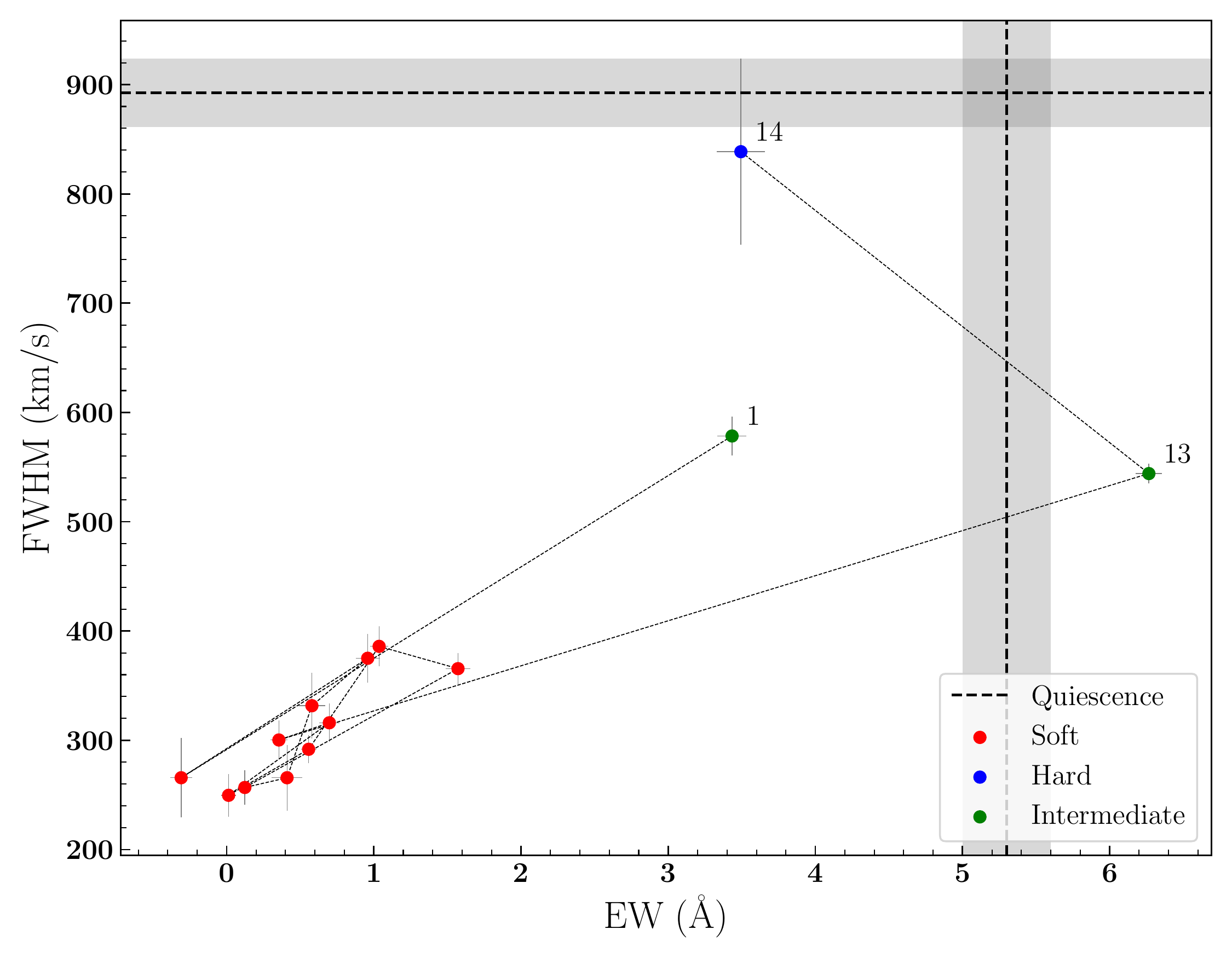}
    \end{subfigure}
    \begin{subfigure}{\columnwidth}
        \includegraphics[width=9truecm]{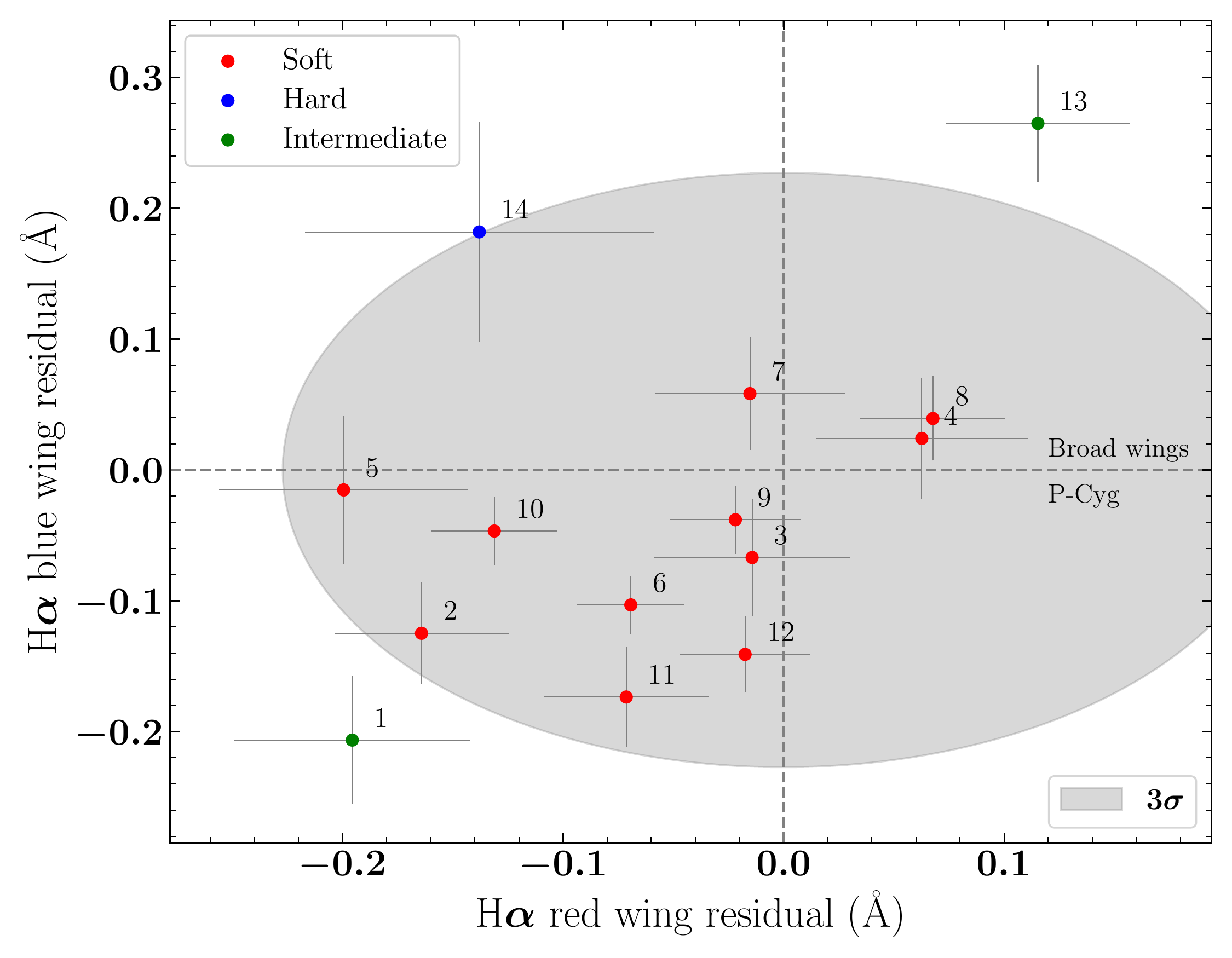}
    \end{subfigure}
    \caption{Diagnostic diagrams. Colour code is the same as in Fig. \ref{fig1_light_HID}. \textit{Left panel:} H$\alpha$ FWHM versus its EW measured in a  44~\mbox{\normalfont\AA} -wide ($\sim$2000~\kms) region centred at the rest frame of the binary. The quiescent values of EW and FWHM (\citealt{Shahbaz1997,Garcia1999}) are indicated as vertical and horizontal black dashed lines, respectively (grey bands indicate uncertainties). The FWHM quiescent value reported in \citet{Garcia1999} (830 $\pm$ 25 \kms) has been downgraded to the spectral resolution of our data. \textit{Right panel}: H$\alpha$ blue and red wing residuals (excesses) for the 14 epochs. The grey-shaded region indicates the 3$\sigma$ significance contour.}
    \label{fig2_opcionB}
\end{figure*}
% --------------------------------------------------------

%%%%%%%%%%%%%%%%%%%%%%%%%%%%%%%%%%%%%%%%%%%%%%%%%%%%%%%%%%%%%%%%%%%%%%
%  RESULTS
%%%%%%%%%%%%%%%%%%%%%%%%%%%%%%%%%%%%%%%%%%%%%%%%%%%%%%%%%%%%%%%%%%%%%%
\section{Analysis and results}\label{cap.results}
The 2016 outburst of Aql~X-1 was first detected by \textit{Swift}/BAT on 29 July (MJD 57598, \citealt{Sanna2016}) and lasted $\sim$~70 days, reaching an X-ray peak luminosity of $\sim$~\num{8.5e37}~\ergs \  (\citealt{LopezNavas2020}, assuming a distance of 4.5 kpc). This translates to $\sim$~0.5 L$_\mathrm{Edd}$ for a 1.4 \Msun \ NS accretor, likely the most luminous outburst ever observed of the source \citep{Gungor2017}. Its X-ray light-curve\footnote{Using monitoring data from the Monitor of All-sky X-ray Image, MAXI (\citealt{Matsuoka2009}).} (upper panel in Fig. \ref{fig1_light_HID}) shows the usual fast rise ($\sim$10 days) followed by a slower decay to quiescence. 
According to their position in the HID\footnote{Using data from MAXI and the Burst Alert Telescope on board the Neil Gehrels Swift Observatory, \textit{Swift}/BAT (\citealt{Barthelmy2005}, \citealt{Krimm2013}).} (bottom panel in Fig. \ref{fig1_light_HID}), most of our observations were obtained during the bright soft state, while the first and the last two epochs correspond to harder states. The X-ray states of epochs \#1 and \#13 are not perfectly determined by this diagram. This difficulty has also been remarked by \citet{DiazTrigo2018}, who classified these observations as hard or intermediate. In this paper, we assume that both epochs, \#1 and \#13, were obtained during the intermediate state, understood as a transitional state between the hard and the soft ones (see e.g. fig. 9 in \citealt{Munoz-Darias2014}).
    
The optical spectra vary as the outburst evolves. They include helium and hydrogen emission lines, such as H$\beta$ (4861~\mbox{\normalfont\AA}), \ion{He}{i} at 5876~\mbox{\normalfont\AA} (\ion{He}{i}–5876), H$\alpha$ (6563~\mbox{\normalfont\AA}), and even \ion{He}{i}--6678 and \ion{He}{i}--7065, which are barely visible only in epoch \#13 (see  Fig. \ref{fig2new}). The Bowen blend and \ion{He}{ii}--4686 are also detected in the R2500V dataset (see \citealt{JimenezIbarra2018} for a detailed analysis of these lines). We studied the presence or absence and profile evolution of the above emission lines, paying special attention to \ion{He}{i}--5876 and H$\alpha$, which are known to be particularly sensitive to optical accretion disc winds (e.g. \citealt{Munoz-Darias2016, Munoz-Darias2020}):

    \ion{He}{i}--5876 is very weak in our dataset, only being detected at the beginning and end of the outburst (not shown). It is double-peaked in epoch~\#1, when the luminosity is rising, but then disappears during the brightest epochs. It is  again present in the last two epochs (\#13 and \#14), when the system was returning to quiescence. However, it is not particularly strong in epoch \#13 (Fig. \ref{fig2new}) and too noisy in \#14. Therefore, we did not include \ion{He}{i}--5876 in our analysis, although we note that no wind signature is apparent.

    H$\alpha$ is the most intense spectral line in our spectra. As commonly observed in LMXBs in outburst (e.g. \citealt{MataSanchez2018}), it shows a single-peaked profile  that changes shape and intensity as the outburst evolves (right panel in Fig. \ref{fig2_evolution}). In the very first epoch, it is broad and intense, becoming weak and narrow at the outburst peak. In epoch~\#13, the line shows its greatest intensity, while a double-peaked profile is likely present in epoch~\#14 as the system returns towards quiescence. This transition from single to double-peaked lines during the return to quiescence (and vice versa when entering the outburst phase) has been observed in numerous LMXBs, although the physical reasons behind this behaviour are not yet clear (see \citealt{Matthews2015} for a discussion on this topic for the case of cataclysmic variables). A detailed analysis of this emission line is presented below.
    
    Although less intense, H$\beta$ is also visible during the outburst as a single-peaked emission line (left panel in Fig. \ref{fig2_evolution}). A broad absorption is present in epoch~\#1, reaching $\sim$~$\pm$1700 \kms. This broad component is not detected during the rest of the outburst, and could be similar to the broad absorptions centred at (approximately) zero velocity observed in helium and Balmer lines in other systems. It has been suggested that these are formed when the accretion disc is optically thick and behaves like a stellar atmosphere \citep{Soria2000, Dubus2001}. As for the case of H$\alpha$, the line is weak at the outburst peak, with epoch \#13 showing the most prominent emission. We note that a redshifted absorption is present in epoch \#10 both in H$\alpha$ and H$\beta$, centred in $\sim$200 \kms. Similar redshifted absorption features, which might suggest the presence of in-falling material (e.g. failed winds), have been observed in other LMXBs and discussed in \citet{Cuneo2020b}. Finally, we note that in the H$\beta$ region this feature is superimposed on a broader absorption, which is visible in every epoch and extends up to $\sim$2000 \kms. This is commonly observed in transients in outburst and could be due to a weak diffuse interstellar band (e.g. \citealt{Buxton2003}).

\subsection{\texorpdfstring{H$\alpha$ analysis}{Ha analysis}}
In order to study the evolution of the H$\alpha$ line we fitted each epoch's profile with a Gaussian function. The offsets of the line with respect to the central wavelength are in the range of -115$\pm$8 to 190$\pm$30 \kms, the latter value corresponding to epoch \#14. The line is also slightly shifted towards the red (93~$\pm$~4~\kms) when it is most intense (epoch \#13). 

\subsubsection{The FWHM-EW diagram}
We plotted the full width at half maximum (FWHM) of H$\alpha$ as a function of its equivalent width (EW, left panel in Fig.~\ref{fig2_opcionB}). This diagram reveals that epochs corresponding to the outburst peak populate a very specific area, with EW \mbox{\textless} 2~\mbox{\normalfont\AA} and FWHM~\mbox{\textless}~400~\kms; both smaller than the observed values during quiescence (EW = 5.3~$\pm$~0.3~\mbox{\normalfont\AA}, \citealt{Shahbaz1997} and FWHM~=~830~$\pm$~ 25~\kms, \citealt{Garcia1999}). Contrastingly, the line is broader (FWHM \mbox{\textgreater} 500~\kms) and closer to the quiescent values in the first and last epochs (i.e. \#1 and \#13--14). In particular, during epoch \#13 we observe the highest EW ( $\sim$ 7~\mbox{\normalfont\AA}). As a result of this evolution, the source draws a loop in the diagram, resembling to some extent the nebular loops observed in V404 Cyg \citep{MataSanchez2018}. 
We note that if we take into account the possible contribution from the interloper (diluting the Aql X-1 emission features; see Sect. \ref{interloper}), epoch \#13 could even become slightly more extreme and epoch \#14 would likely approach the quiescent value (as is the case for the FWHM).
        
\subsubsection{The excesses diagram} \label{SectExcessesDiagram} 
We performed a systematic search for outflow signatures using the H$\alpha$ excesses diagram presented in \citet{MataSanchez2018}. In a first step, we masked the emission line wings and fit this with a Gaussian. For this fit, we left the FWHM, high, and centre of the Gaussian free to vary. Subsequently, the Gaussian fit is subtracted from the spectral line maintaining a normalised continuum. Finally, the EW of the resulting residuals is measured in the red and blue wings previously masked during the Gaussian fit. These are the so-called red and blue excesses, EW$_\mathrm{r+,}$ and EW$_\mathrm{b-}$, respectively. The regions where EW$_\mathrm{r+}$ and EW$_\mathrm{b-}$ are measured must be within the masked ranges, with specific widths and positions depending on the source. In this particular case, we used 500 to 1000~\kms \  and -1000 to -500~\kms \ for both masking and measuring the excesses (see Sect. 3.2.4 and fig. 11 in \citealt{MataSanchez2018} for further details on this technique). Significance levels were computed following \citet{Munoz-Darias2019}. In particular, we measured the EW of the continuum in a few nearby regions using masks of the same width than for EW$_\mathrm{r+}$ and EW$_\mathrm{b-}$. These continuum residuals show a Gaussian distribution that can be fitted in order to derive a sigma value, which is subsequently used to trace significance contours. We computed these residuals in six regions between~$\pm$~2000 and~$\pm$~4000 \kms. 

The resulting diagram is presented in the right panel of Fig. \ref{fig2_opcionB}. Epochs \#1 and \#13 are the only ones with significant excesses (\mbox{\textgreater} 3$\sigma$). Epoch \#1 sits in the region of negative excesses. This is consistent with the presence of a weak underlying absorption, such as that clearly visible in H$\beta$ in the very same observation (see Fig. \ref{fig2_evolution}). We note that Balmer broad absorptions of varying intensity are not uncommon in LMXBs (e.g. \citealt{Rahoui2014,Jimenez-Ibarra2019a}). In addition, the red wing of H$\beta$ can be also affected by an additional, underlying absorption (present in every epoch) likely related to a weak diffuse interstellar absorption (DIB) (see e.g. \citealt{Kaur2012,Cuneo2020b} for a discussion). The excesses in epoch \#13 are consistent with the presence of a broad emission component and, in fact, a two-Gaussian model significantly improves the fit, reducing \qhired\ from \textasciitilde8 to \textasciitilde2 (see Fig. \ref{fig3_haEp13_heiEp13}). A two-Gaussian model also provided a better fit to the H$\alpha$ profiles in V404 Cyg during the nebular phase (see Sect. 4 and fig. 2 in \citealt{Casares2019}). The full width at zero intensity (4.29$\sigma$) of the (additional) broad Gaussian is 810~$\pm$~170~\kms. 
As we discuss above, the derived EW values (and therefore the excesses) could be lower than the actual ones due to the contamination by the interloper when the source approaches quiescence. This is particularly true for epoch \#14, which also has the largest errors. Hence, we do not derive any conclusion from this epoch. 

\subsection{Balmer decrement}
\begin{figure}
\centering
\begin{subfigure}{\columnwidth}
    \centering
    \includegraphics[width=9truecm]{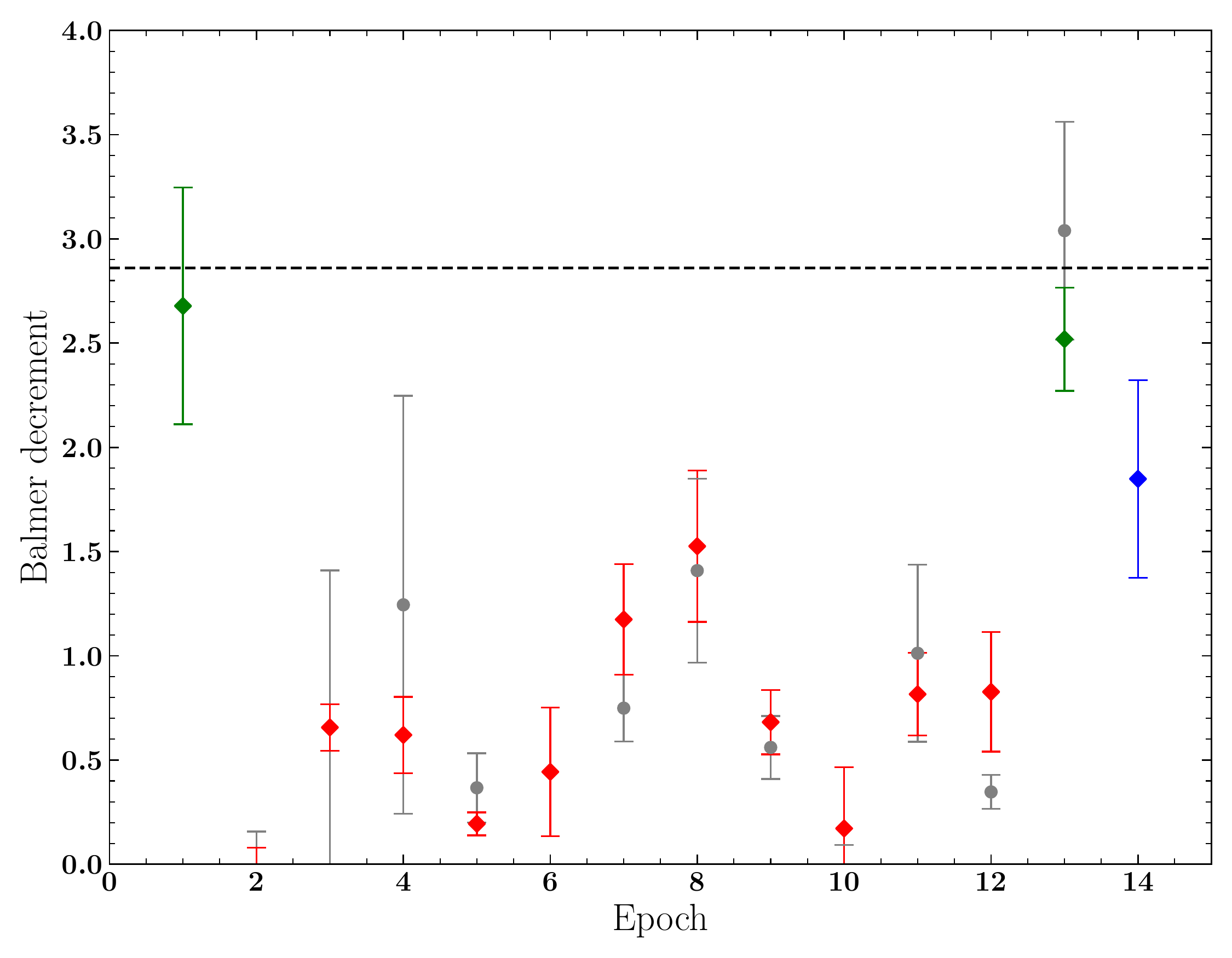}
\end{subfigure}
\caption{Evolution of the Balmer decrement.  Grey dots were computed by estimating the H$\beta$ flux as twice the blue half flux. Coloured dots (following the colour code in Fig. \ref{fig1_light_HID}) were derived by adding a Gaussian component to the continuum fitting process (see text). The black dashed line indicates case B recombination (\citealt{Osterbrock1989}).}
\label{fig_BD}
\end{figure}

The H$\alpha$-to-H$\beta$ flux ratio, known as the Balmer decrement (BD), is a good indicator of the physical properties of nebulae \citep{Baker1938} and accretion discs \citep{WilliamsShipman1988}. In particular, it is known that neutral hydrogen self-absorption can increase this ratio, especially in relatively low-density conditions (e.g. \citealt{Drake1980}), such as those of nova (e.g. \citealt{Iijima2003}) and X-ray binary (e.g. \citealt{Munoz-Darias2016}) ejections. Therefore, the BD can be used as an additional indicator for the presence of nebulosities. We calculated the BD using the flux calibrated R1000B data. In a first step and in order to avoid the broad absorption visible in the red wing of H$\beta$, we computed its flux as twice the blue half flux (-1000 to 0 \kms), while -1000 to 1000 \kms \ was used for H$\alpha$ (grey dots in Fig.~\ref{fig_BD}). The results are consistent with those obtained by using the total H$\beta$ flux, which have larger errors. However, this method could not be applied to epochs \#1 and \#3 since they are dominated by a very broad absorption that also affects the blue wing (see Fig.~\ref{fig2_evolution}). Likewise, results from epochs \#2, \#6, \#10, and \#14 were also excluded given that H$\beta$ is weak and affected by absorption components, which resulted in very large errors.
In a second step, we computed the BD for every epoch by adding a Gaussian component (accounting for the broad absorption in H$\beta$)  in the continuum fitting process (see coloured dots in Fig.~\ref{fig_BD}). For the epochs in which both methods could be applied, the results are consistent within $1.5\sigma$ in all the cases. We find low BD values ($\lesssim$ 1.5) for every soft state epoch. Contrastingly, epoch \#13 (see Fig.~\ref{fig_BD}) is characterised by a BD of $\sim 3$. This value could be up to 20\% lower if we consider the 15\% contribution of the interloper to the total flux (adopting EW$\sim 1.6$ \AA \ for the underlying absorption features of H$\beta$ and H$\alpha$ ; \citealt{Joner2015}). Therefore, the BD for epoch \#13 is roughly consistent with the canonical value for case B recombination (BD=2.86 for T=10$^4$ K and N$_e$=10$^2$ cm$^{-3}$; \citealt{Osterbrock1989}), which represents an opaque (i.e. optically thick) nebula to ionising radiation \citep{Baker1938}. 
This might be also the case for epoch \#1, although we note that the obtained value can be very sensitive to the significant broad absorptions present in the Balmer lines during this epoch. 

Finally, we note that given that the flux measurements are not obtained by directly comparing the target's spectrum with that of field stars (see Sect. \ref{observations}), they might be affected by systematics in the calibration process (e.g. wavelength-dependent slit losses). However,  BD values significantly lower than that of epoch \#13 are typically obtained (Fig. \ref{fig_BD}), and this epoch is also characterized by other observables (see above) consistent with the higher BD value measured (Fig. \ref{fig_BD}).
    
% ------------------------ Figure 5 --------------------------
% --------------------------------------------------------

\subsection{Ionisation state}
In some cases, the visibility of cold winds has been found to be affected by the ionisation state of the gas, which is preferentially observed at low ionisation (\citealt{Munoz-Darias2016, Munoz-Darias2019}). Since the disc is highly irradiated during the outburst, its ionisation state can be traced by the intensity of the spectral lines with the higher ionisation potentials. 
In the context of optical spectra of LMXBs, the flux ratio 
$I_{\mathrm{ratio}}=\ion{He}{ii}~(4686~\mbox{\normalfont\AA})/$H$\beta $ can be used as a good indicator, since it is hardly affected by flux calibration issues given that these two lines are only $\sim$200~$\AA$ apart. As observed in other systems \citep[e.g.,][]{MataSanchez2018}, epochs corresponding to the outburst peak show the highest $I_{\mathrm{ratio}}$ values (>1.7). The lowest values correspond to the final epochs, in which the ratio decreases to 1.25$\pm$0.25 (epoch \#13). This temporal evolution is consistent with the trend observed in the EW of the Bowen blend and \ion{He}{ii}--4686 (B+\ion{He}{ii}), which can also be used as a tracer for the ionisation of the disc. The highest values are found at the outburst peak B+\ion{He}{ii} ($\sim$~4--5), while the lowest correspond to epochs \#1 and \#13 ($\sim$ 3.25--3.6). We note that the deep P-Cyg profiles observed in V404 Cyg are characterised by $I_{\mathrm{ratio}} \lesssim 1$.

%%%%%%%%%%%%%%%%%%%%%%%%%%%%%%%%%%%%%%%%%%%%%%%%%%%%%%%%%%%%%%%%%%%%%%
% DISCUSSION
%%%%%%%%%%%%%%%%%%%%%%%%%%%%%%%%%%%%%%%%%%%%%%%%%%%%%%%%%%%%%%%%%%%%%
\section{Discussion}
We have presented optical spectroscopy of Aql~X-1 obtained with the GTC-10.4m telescope during its 2016 outburst, the brightest in recent years, which lasted $\sim$~70 days. We studied the evolution of the most important spectral lines and searched for spectral features that are commonly associated with accretion disc winds. Although Aql~X-1 is one of the most studied LMXBs, this is the first time that this kind of analysis has been performed in this system. 
    
\begin{figure}
\centering
\begin{subfigure}{\columnwidth}
    \centering
    \includegraphics[width=9truecm]{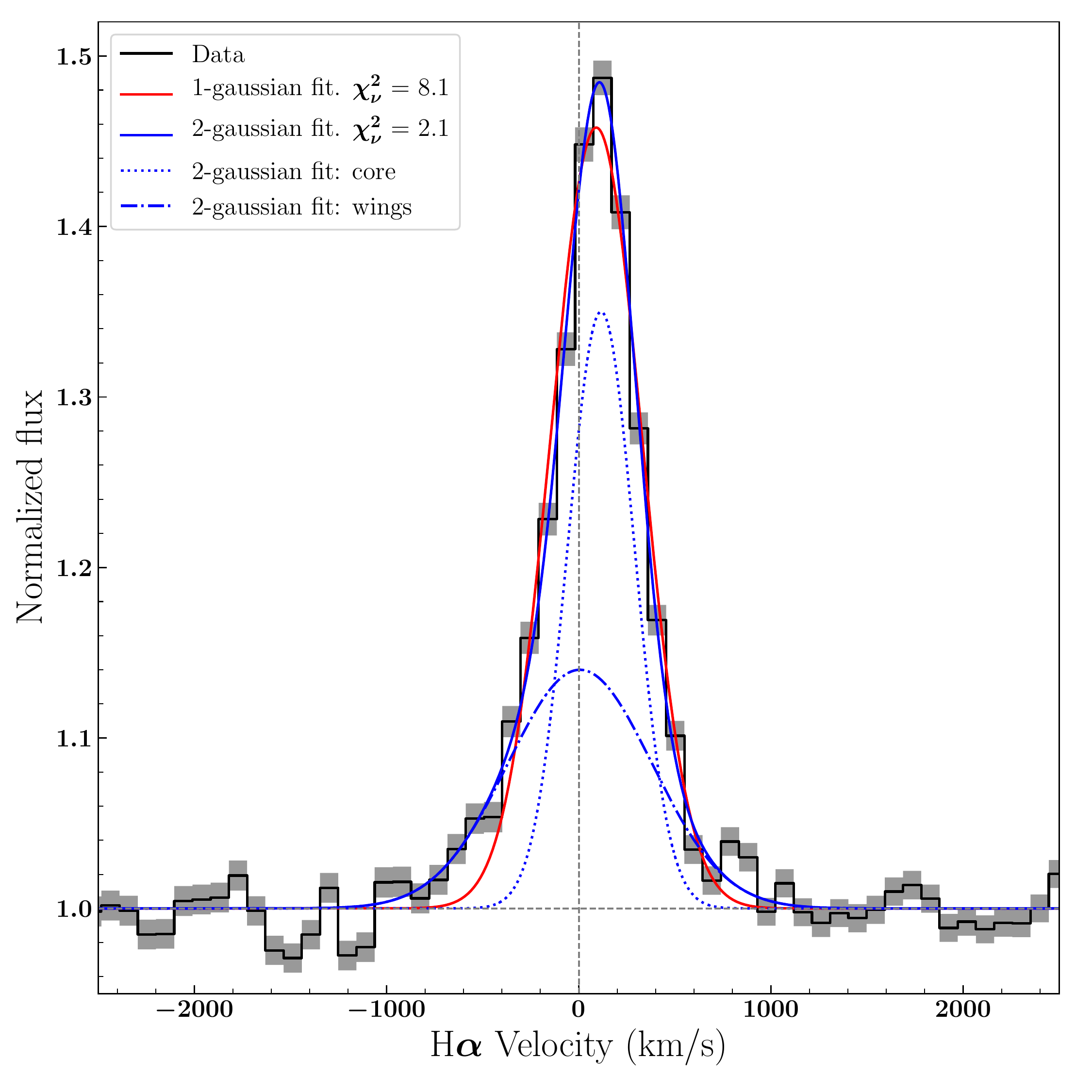}
\end{subfigure}
\caption{H$\alpha$ in epoch \#13. The shadowing indicates the error in normalised flux. Red and blue solid lines represent the Gaussian and the two-Gaussian fits, respectively. Blue dotted and dash-dotted lines correspond to the two components of the latter. 
}
\label{fig3_haEp13_heiEp13}
\end{figure}

While the \ion{He}{i}--5876 emission line (arguably the most effective optical wind tracer) is too weak for a detailed analysis, the H$\alpha$ excesses diagram reveals positive excesses in one of the final epochs (\#13), both in the red and blue halves. This indicates that the line cannot be described by a single Gaussian profile. Instead, it needs an additional broad, low-intensity component (Fig. \ref{fig3_haEp13_heiEp13} and Sect. \ref{SectExcessesDiagram}), which could be interpreted as the signature of a nebular phase that occurred at the end of the outburst, similar to those detected by the same method in V404 Cyg (\citealt{MataSanchez2018}). The bulk velocity of the ejecta would be $\sim$~800~\kms, which is consistent with typical values of X-ray winds in BH transients (\citealt{Ponti2016}).  

% ------------------------ Figure 6 --------------------------
% --------------------------------------------------------      

This interpretation is supported, albeit indirectly, by the FWHM--EW diagram. On the one hand, the highest value of the H$\alpha$ EW is observed during this epoch. This implies that during the final drop in brightness, with a decrease of over 50\% in the X-ray luminosity within the eight days preceding epoch \#13 (Fig. \ref{fig1_light_HID}), the optical continuum (likely arising in the disc) is decaying faster than the emission line flux, which could come from previously launched ejecta while cooling down and recombining. On the other hand, the source describes a loop in the diagram as the outburst evolves; an observational pattern that has been found to be linked to the presence of nebular phases in V404 Cyg \citep{MataSanchez2018}. We note that the anticlockwise loop pattern present in the diagram is different from the clockwise evolution found in V404 Cyg (fig. 9 in \citealt{MataSanchez2018}). This is due to the conspicuous wind signatures present in V404 Cyg during the whole outburst. This made the FWHM of the Balmer lines larger than the quiescent value, while the opposite is commonly observed in LMXBs  (e.g. \citealt{Munoz-Darias2013b, Casares2015}), as is the case for Aql X-1.

The evolution of the BD does not reveal extreme values, as those observed in V404 Cyg during some epochs (BD $\sim$ 6), but it is consistent with the above scenario. During the epochs corresponding to the most luminous phases of the outburst, the BD is close to unity (see Fig. \ref{fig_BD}), which is consistent with an optically thick, irradiated accretion disc, as is found in cataclysmic variables \citep{Drake1980, Williams1980, WilliamsShipman1988, Tomsick2016}. However, epoch \#13 shows a significantly higher value, BD~$\sim 3$, which is in agreement with the case B recombination in a low-density nebula (\citealt{Osterbrock1989}). As a matter of fact, high BD values ($\gtrsim3$) were typically observed in the 2015 outburst of V404 Cyg concurrently with P-Cyg profiles in hydrogen and helium emission lines \citep{Munoz-Darias2016, MataSanchez2018}.
        
\subsection{Signatures of optical outflows in previous outbursts}
This is the first time that an analysis of this type has been performed in Aql~X-1 using a large database covering almost the entire outburst. However, the intense and broad H$\alpha$ profile observed in epoch \#13 resembles (at least in strength) that witnessed by \citet{Shahbaz1998} during the early phases of the August 1997 outburst \citep{Charles1997, Chevalier1997}. In addition, the authors reported a possible P-Cyg profile in H$\beta$, although a normalised spectrum was not provided. This might suggest that an optical outflow was present in this event, which was significantly less luminous than the one in 2016 (\citealt{Gungor2017} estimated its X-ray peak luminosity to be $\sim$~\num{2.3e37}~\ergs ; i.e. $\sim$26\% of the 2016 outburst peak). However, it is important to bear in mind that LMXBs are known to show broad absorptions in He and Balmer lines centred at (approximately) zero velocity during some stages of the outburst (see Sect. \ref{cap.results} and Fig. \ref{fig2_evolution}). Finally, we note that \citet{Cornelisse2007} reported on the presence of stationary narrow emission lines (\ion{N}{iii} and \ion{He}{ii}--4686) during the rising and decaying stages of the 2004 outburst (i.e. not at the outburst peak). The nature of these lines is unclear, but they may arise in stationary material surrounding the binary.

\subsection{Comparison with other systems and alternative scenarios} 
Our observations of Aql~X-1 during its luminous 2016 outburst ($\sim$~0.5 L$_{\mathrm{Edd}}$) have revealed the presence of strong and broad H$\alpha$ emission during its final stages. This might be associated with the presence of a nebular phase produced by ejecta moving at $\sim$~800 \kms. Broad emission wings in Balmer lines were also seen during the luminous outbursts of the BH transients V404 Cyg and V4641 Sgr. However, those were (presumably) super-Eddington outbursts, and the nebular phases showed higher terminal velocities ($\sim$~3000 \kms) and line intensities. Nevertheless, the velocity derived from the breadth of the H$\alpha$ line in epoch \# 13 ($\sim$~800 \kms) is within the usual range found in cold winds in LMXBs (a few hundreds to 3000 \kms).

If the intense H$\alpha$ emission is associated with an outflow, Aql~X-1 would become the second NS transient, after Swift~J1858.6-0814, where signatures of optical outflows are reported. The latter showed conspicuous P-Cyg profiles during its atypical 2018--2019 outburst. These were found during relatively (X-ray) hard phases characterised by variable radio jet emission (\citealt{Munoz-Darias2020, vandenEijnden2020}; see also \citealt{Buisson2020} for a possible X-ray wind detection). In this regard, we note that although our first epoch was taken during the rising phase of the outburst (see Fig. \ref{fig1_light_HID}), the initial hard state was missed. This phase is typically shorter in Aql~X-1 (and in other NS transients) than for BH transients (see \citealt{Munoz-Darias2014}). In addition, we note that there are significant differences in the orbital inclinations  of the above systems. While it is low-to-intermediate for Aql~X-1 ($23^\circ$<\textit{i}<$53^\circ$ in the most conservative scenario of \citealt{MataSanchez2017}), Swift~J1858.6-0814 has shown high inclination features (\citealt{Buisson2020b}). A high orbital inclination seems to favour the detection  of optical outflows (see \citealt{Munoz-Darias2020}) and X-ray winds (\citealt{Ponti2012}, \citealt{DiazTrigo2016}).  Our results suggest that nebular phases (e.g. strong emission lines with extended wings) might offer the possibility of detecting winds in lower inclination LMXBs, such as Aql~X-1, where P-Cyg profiles (i.e. blue-shifted absorptions) might be harder to detect  (see also \citealt{Higginbottom2019} for theoretical studies). This is also supported by the observation of a P$\beta$ broad emission component in the BH transient GX 339--4 during its 2010 outburst (\citealt{Rahoui2014}). It was detected during two observations taken during the initial hard state of the outburst, and it was suggested to arise in an extended envelope covering the inner accretion disc, which could also indicate the presence of an accretion disc wind. As it happens for epoch \#13 in our study, the BDs reported for these observations ($\sim 1.5$ and 2.5) are significantly higher than that of the soft state epoch ($\sim 1$) analysed in \citet{Rahoui2014}.  GX~339-4 has not displayed high inclination features and is thought to be seen through an intermediate line of sight (see \citealt{Munoz-Darias2013,Heida2017}). 
  
An alternative explanation for the H$\alpha$ emission observed in epoch \#13 could be the presence of a lower density accretion disc atmosphere, which would be extensively irradiated during the brightest phases of the outburst (i.e. soft state). In this scenario, the optically thin emitting gas would be bound to the system and the breath of the H$\alpha$ component simply due to velocity dispersion. However, the relatively low inclination value of the system ($i\lesssim 50^{\circ}$) does not seem to favour this interpretation, unless the size of this atmosphere is remarkably large. Interestingly, peculiar X-ray dips were observed in the 2011 outburst of the Aql~X-1 \citep{Galloway2016}. By contrast to the typical behaviour of regular dippers, these dips are 'intermittent' and have been reported in only two observations out of the $\sim$ 500 performed by the \textit{Rossi X-ray Timing Explorer} over more than 15 years and across different outbursts. The dips were observed in data taken during the faint hard state in one case and during (or just before) the hard-to-soft transition in the other. The exact origin of these transient features is unknown, but they could be (very tentatively) related with 'clouds' of material (e.g. ejecta) or indicate that the outer disc significantly thickens during some phases of the outburst (particularly during the hard state). It is worth noting that the latter scenario might also be connected to the launch of optical outflows, as is likely the case of the BH optical dipper Swift J1357.2-0933 (\citealt{JimenezIbarra2019b, Charles2019}; see also \citealt{Corral-Santana2013}).

%%%%%%%%%%%%%%%%%%%%%%%%%%%%%%%%%%%%%%%%%%%%%%%%%%%%%%%%%%%%%%%%%%%%%%
% CONCLUSIONS
%%%%%%%%%%%%%%%%%%%%%%%%%%%%%%%%%%%%%%%%%%%%%%%%%%%%%%%%%%%%%%%%%%%%%%
\section{Conclusions}
We present a detailed optical spectroscopic monitoring of Aql~X-1 during its 2016 outburst, the most luminous in recent times, focussed on the evolution of emission lines that are known to be good tracers of accretion disc winds. We do not detect\ red disc wind signatures, such as P-Cyg profiles. However, the properties of the strong H$\alpha$ broad emission component witnessed during the decaying phase of the outburst indicates the presence of optically thin emitting material. This can be interpreted as a nebular phase produced by previously launched ejecta, similar to those already detected in a handful of BH transients. Additional spectroscopic follow-up of this system during forthcoming accretion events (also covering the very early stages of the outbursts missed here) should be able to confirm the presence of optical outflows in Aql~X-1 and to determine their main observational properties. 

\begin{acknowledgements}
We are thankful to the anonymous referee for constructive comments that have improved this paper. We  acknowledge support from the State Research Agency (AEI) of the Spanish Ministry of Science, Innovation and Universities (MCIU) and the European Regional Development Fund (ERDF) under grant AYA2017- 83216-P. TMD acknowledges support from the Consejer\'ia de Econom\'ia, Conocimiento y Empleo del Gobierno de Canarias and the ERDF under grant with reference ProID2020 010104. TMD acknowledges support via the Ramón y Cajal Fellowship RYC-2015-18148. DMS acknowledges support from the ERC under the European Union’s Horizon 2020 research and innovation programme (grant agreement no. 715051; Spiders).
Based on observations made with the Gran Telescopio Canarias (GTC), installed at the Spanish Observatorio del Roque de los Muchachos of the Instituto de Astrofísica de Canarias, in the island of La Palma. This research has made use of MAXI data provided by RIKEN, JAXA and the MAXI team. We acknowledge the use of public data from the \textit{Swift} data archive. \textsc{Molly} software developed by Tom Marsh is gratefully acknowledged. 
\end{acknowledgements}

%%%%%%%%%%%%%%%%%%%%%%%%%%%%%%%%%%%%%%%%%%%%%%%%%%%%%%%%%%%%%%%%%%%%%%
% BIBLIOGRAPHY
%%%%%%%%%%%%%%%%%%%%%%%%%%%%%%%%%%%%%%%%%%%%%%%%%%%%%%%%%%%%%%%%%%%%%%
\bibpunct{(}{)}{;}{a}{}{,}
\bibliographystyle{aa}

\end{document}